\documentclass[a4paper]{jpconf}
\usepackage{graphicx}
\usepackage{amsmath}
\usepackage{citesort}
\usepackage{csquotes}

\begin{document}
\title{ Some aspects of the canonical analysis of Reuter-Weyer RG improved Einstein-Hilbert action}
\author{Gabriele Gionti, S.J.}

\address{Specola Vaticana, V-00120 Vatican City, Vatican City State\\ and
Vatican Observatory Research Group Steward Observatory, The University Of Arizona,\\ 
933 North Cherry Avenue, Tucson, Arizona 85721, USA.\\
INFN, Laboratori Nazionali di Frascati, Via E. Fermi 40, 00044 Frascati, Italy.\\}

\ead{ggionti@specola.va}

\begin{abstract} 
A canonical analysis of RG improved action of the Einstein-Hilbert functional is performed. The gravitational and cosmological constants as function of the space-time coordinates are treated as external non-geometrical fields. Dirac's constraint analysis is performed, in the general case, up to secondary constraints. The constraints are second class and, in general, the problem appears to be technically complicated. This fact suggests studying the Dirac's constraint analysis of the related Brans-Dicke theory. It exhibits a Dirac's constraint algebra similar to Einstein's geometrodynamics except that the Poisson Brackets between Hamiltonian-Hamiltonian constraints is not only linear combination of the momentum constraints but also of a term note reducible to linear combination of the constraint and proportional to the extrinsic curvature. This shows that Branse-Dicke geometrodynamics is inequivalent to Einstein General Relativity geometrodynamics.

\noindent A simplified FLRW minisuperspace model based on the RG improved Einstein Hilbert action contains Bouncing and Emergent Universes for values of $K=-1, 0, 1$

\end{abstract}

\section{Introduction}

	Einstein General Relativity appears to be a successful phenomenological theory at laboratory, solar system, galactic and in general at distances bigger than the Planck length $l>>l_{Pl}\equiv \frac{1}{\sqrt{G}}\approx10^{-33}cm.$. All the classical tests, precession of Mercury, bending of the light rays close to massive bodies \cite{Weinberg1972}, and the recent detection of the Gravitational Waves \cite{TestGRLIGO2016}, attest Einstein General Relativity is a sound classical theory. But as it is even known at the popular level \cite{Hawking1973}, Einstein General Relativity has an initial singularity. This fact means that General Relativity is no longer predictive around the singularity. People have speculated that this breakdown of the physical laws signals the emergence of a new physics. Although the previous statement is still matter of debate, certainly it is quite well known that matter at atomic and subatomic level behaves according to the laws of Quantum Mechanics. The attempt to formulate a sound Quantum Theory of Gravity (Quantum Gravity) is as yet unrealized. There are many different attempts: String Theory, Loop Quantum Gravity, Non-Commutative Geometry, Dynamical Triangulation and  Causal Dynamical Triangulations, Asymptotic Safety etc.  None of these approaches have come yet to a final theory of Quantum Gravity which satisfies the expectations of the entire scientific community.

Quantized General Relativity, along the guidelines of Quantum Field theory, is perturbatively non-renormalizable. The Newton constant $G$ has the dimension of the inverse of a square length. Therefore, we have to add a number of counter terms which increase as the loop orders do. The renormalizzation process introduces infinitely many parameters so that the resulting theory does not have any predictive power \cite{lauscherreuter2001}. In general a theory is considered "fundamental" if it is perturbatively renormalizable. This means that its infinities can be absorbed by redefining a finite number of parameters. It follows that Quantum General Relativity is not a fundamental theory in this sense. But this is not the end of the story, because there exist fundamental theories which are non-perturbatively renormalizable. This non-perturbative renormalizability, introduced by K. Wilson  \cite{wilsonfourQFT}, is related to the existence of a Non-Gaussian Fixed Point (NGFP) in the space of the parameters which guarantee the finiteness of the theory in the ultraviolet limit\cite{NiederReuter}.

Stephen Weinberg \cite{1979weinberg} proposed the Asymptotic Safety conjecture. He suggested that Einstein General Relativity might be defined non-perturbatively at the non-Gaussian fixed point . He himself proved that NGFP exists in 2+$\epsilon$ dimensions \cite{1979weinberg}. In d=4 there has been no progress because of the lack of a calculation scheme.  Recently, \cite{Reuter1996}, progress has been made using the \textquotedblleft effective average action\textquotedblright. Implementing the \textquotedblleft Einstein  Hilbert Truncation\textquotedblright \cite{2001ReuterSaueressig}, it has been shown that there is a NGFP. There is strong evidence that the fixed point exists in the exact theory as well.    

\section{Renormalization Group approach}

A Wilson-type, coarse-grained, free energy functional $\Gamma_{k}\left[g_{\mu\nu}\right]$ is defined in the following way: $\Gamma_{k}\left[g_{\mu\nu}\right]$
contains all the quantum fluctuations with momenta $p>k$ and not yet of those with $p<k$ \cite{Reuter2007}. The modes $p<k$ are suppressed in the path-integral by a mass-square  type term $R_{k}(p^{2})$.
The behavior of the free-energy functional interpolates between $\Gamma_{k\mapsto \infty}=S$, $S$ being the classical (bare) action, and $\Gamma_{k\mapsto 0}=\Gamma$, $\Gamma$ being the standard effective action. $\Gamma_{k}$ satisfies the RG-equation, called also the Wetterich equation \cite{WetterichFRG}, 

\begin{equation}
k\partial_{k}\Gamma_{k}=\frac{1}{2}Tr\left[(\delta^2\Gamma_k + R_k)^{-1}k\partial_k R_k\right]
\label{Wett}
\end{equation}

In general, since this $RG$-equation is very complicated, one adopts a powerlul non perturbative approximation scheme: truncates the space of the action functional and projects the RG flow onto a finite dimensional space. That is to say, one considers that the free energy functional $\Gamma_{k}$, formally, can be expanded in the following way

\begin{equation}
\Gamma_{k}[\cdot]=\sum_{i=0}^{N}g_{i}(k)k^{d_i}I_i[\cdot]\;\;\;\;,
\label{svilop}
\end{equation}

where $I_{i}[\cdot]$ are given local or non local functionals" of the fields and a-dimensional coefficients $g_{i}(k)$. 
In the case of gravity, the following truncation antsatz is usually made:

\begin{equation}
I_{0}[g]=\int d^{4}x\sqrt{g}\;\;\;\;,I_{1}[g]=\int d^{4}x\sqrt{g}R\;\;\;\;,I_{2}[g]=\int d^{4}x\sqrt{g}R^{2}\;\;\;\;,\mathrm{etc.}
\label{trunca}
\end{equation}

The simplest truncation is the Einstein-Hilbert truncation which looks like

\begin{equation}
\Gamma_{k}=-\frac{1}{16\pi G_{k}} \int d^{4}x  \left( R-2 \bar{\lambda}_{k} \right) +\mathrm{g.f. } +\mathrm{g.t.}\;\;\;\,, 
\label{truncaEH}
\end{equation}

here g.f. are classical gauge fixing terms, while g.h. are ghost terms. There are two running parameters $G_{k}$, the Newton constant, which can be written in a dimensionless way as   $g(k)=k^{2}G_{k}$. In the same manner, the cosmological constant $\bar{\lambda}_{k}$ becomes $\lambda(k)=\bar{\lambda}_{k}/k^{2}$.

Inserting this ansatz into the flow (Wetterich) equation, one obtains "a projection" onto e finite dimensional space  \cite{NiederReuter}
\begin{equation}
Tr[...]=(...)\int \sqrt{g} +(...)\int \sqrt{g}R +...\;\;\;\;, 
\label{expo}
\end{equation}

and then the following finite-dimensional RG equations 

\begin{eqnarray}
k\partial_{k}g(k)=\beta_{g}(g,\lambda)\\ \nonumber
k\partial_{k}\lambda(k)=\beta_{\lambda}(g,\lambda)\;.
\label{finite}
\end{eqnarray}

The solutions of these equations provide the scaling relation for the a-dimensional gravitational constant $g(k)$ and the a-dimensional cosmological constant $\lambda(k)$. 

A point $({g}_{\star}, {\lambda}_{\star})$ is a NGFP if it is a non trivial zero of the beta-functions, 
that is $\beta_{g}({g}_{\star}, {\lambda}_{\star})=0 \;  \beta_{\lambda}({g}_{\star}, \lambda_{\star})=0$ and $({g}_{\star}, {\lambda}_{\star})\neq 0$. 

\section{Reuter-Weyer action proposal and its Hamiltonian}

Reuter and Weyer $\cite{ReuterWeyer2004}$ proposed a modified Einstein-Hilbert action with a non geometrical field $G(x)$ and $\Lambda(x)$. 

\begin{equation}
S_{mEH}[g,G(x),\Lambda(x)]\equiv \frac{1}{16\pi}\int d^{4}x\sqrt{-g}\left(\frac{R}{G(x)}-2\frac{\Lambda(x)}{G(x)}\right)\;\;\;\;.
\label{mEH}
\end{equation}

They made the hypothesis that the functional form of the gravitational constant, as function of the Space-Time coordinates, and of the cosmological constant are determined completely by the Renormalization Group and  are independent of the metric tensor $g$. In other words, the variation of the metric tensor $g$ in the action functional $S_{mEH}$ does not affect $G(x)$ and $\Lambda(x)$. 

Looking carefully to the action (\ref{mEH}) and positing $\phi(x)=\frac{1}{G(x)}$, it looks like a Brans-Dicke\cite{BransDicke} theory without the kinetic term for the scalar field $\phi(x)$. Reference $\cite{ReuterWeyer2004}$ discusses the fact that the equation of motion of this theory should impose integrability condition on $G(x)$ and $\Lambda(x)$. Since the functional form of $G(x)$ and $ \Lambda(x)$ is fixed once we make the cut-off identification $k(x)$, the equation of motion finally will put constraints on the cut-off identification. 

We want to study the Hamiltonian Theory (see also \cite{Olmo} for a parallel study) derived by this action.   

The first step of this process is to consider a split of the Space-Time $(M,g)$ in which the manifold M becomes ,topologically,  $M=R\times \Sigma$: $R$ is a one dimensional space, the time direction, $\Sigma$ is a three dimensional space-like surface embedded in $M$,and $g$ becomes the so called ADM \cite{ArnowittDeserMisner}  metric

\begin{equation}
g=-(N^{2}-N_{i}N^{i})dt \otimes dt +N_{i}(dx^{i} \otimes dt
+dt \otimes dx^{i})+h_{ij}dx^{i} \otimes dx^{j}\;\;\;\;, 
\label{metricADM}
\end{equation}

$N=N(t,x)$ is the so called lapse function and $N^{i}=N^{i}(t,x)$ are the shift functions.

 The York-boundary term \cite{1986York} is introduced in the action \eqref{mEH} to make it a differential functional under the variation $\delta g$ af the metric $g$, then, implementing the ADM metric, the action functional becomes
 
 \begin{equation}
S_{ADM}[h_{ij},N, N^{i}]=\frac{1}{16\pi}\int_{R \times \Sigma}dt d^{3}x \sqrt{h}N \frac{1}{G(t,x)}\left({}^{4}R-2 \Lambda(t,x)\right)+\frac{1}{8\pi} \int_{\partial M}{K \sqrt{h}\over G(t,x)}d^{3}x\;\;.
\label{ADMnormal}
\end{equation}

Now we introduce some identities as  in \cite{BonannoEspositoRubano}

\begin{eqnarray}
{1 \over G}\left(K\sqrt{h}\right),_{0}&=&{G,_{0} \over G^{2}}K\sqrt{h}+\left(K\sqrt{h} \over G \right)_{,0} \label{idento1}\\ 
\
{1 \over G}{\partial f^{i} \over \partial x^{i}}&=&{G_{,i} \over G^{2}}f^{i} + {\partial \over \partial x^{i}} \left(f^{i} \over G \right)\;\;.
\label{idento}
\end{eqnarray}

Here $K$ is the trace, performed via the three dimensional metric $h_{ij}$ on the three surface $\Sigma$, of the extrinsic curvature tensor

\begin{equation}
K_{ij}=\frac{1}{2}(-\frac{\partial h_{ij}}{\partial t}+{\bar \nabla}_{i}N_{j}+{\bar \nabla}_{j}N_{i})
\label{curvextrin}
\end{equation}

${\bar \nabla}_{i}N_{j}$ is the covariant derivative on the space-like surface defined through the three-dimensional metric $h_{ij}$   and $f^{i}$ is a vector function defined in the following way \cite{DeWittI1967} \cite{BonannoEspositoRubano}

\begin{equation}
f^{i}\equiv \sqrt{h} \left(K N^{i} - h^{ij} N,_{j}\right)\;\;\;\;. 
\label{bondo}
\end{equation}

The ADM Lagrangian density ${\cal L}_{ADM}$ is then (for all details see \cite{GiontiCorfu2017})

\begin{equation}
 {\cal L}_{ADM}\equiv {1\over 16\pi} \left[{N \sqrt{h}\over G}(K_{ij}K^{ij}
-K^{2}+{ }^{(3)}R-2 \Lambda)
-2{G_{,0}\over G^{2}}K \sqrt{h}
+2{G_{,i}f^{i}\over G^{2}}\right]\;\;\;\;.
\label{densiLagr}
\end{equation}

from this Lagrangian density, we can compute the spatial momentum $\pi^{ij}$ and get

\begin{equation}
{\pi}^{ij}={\partial{{\cal L}_{ADM}} \over \partial{\dot h}_{ij}}= 
-\frac{\sqrt h}{16\pi G}\left(K^{ij}-h^{ij}K\right) +\frac{{\sqrt h}\;h^{ij}}{16\pi N G^{2}}\left(G_{,0}-G_{,k}N^{k}\right)\;\;\;\;. 
\label{momenta}
\end{equation}

In a straight-forward, \cite{GiontiCorfu2017}, it is possible to see that the following re-definition of the spatial momenta 

\begin{equation}
{\tilde \pi}^{ij}=\pi^{ij}-\frac{{\sqrt h}\;h^{ij}}{16\pi N G^{2}}\left(G_{,0}-G_{,k}N^{k}\right)
=-\frac{\sqrt h}{16\pi G}\left(K^{ij}-h^{ij}K\right)
\label{newmomnta}
\end{equation}

allows the definition of the following transformation of coordinates 

\begin{equation}
\left (N, N^{i}, h_{ij},\pi, \pi_{i}, \pi^{ij}\right)\mapsto \left (N, N^{i}, h_{ij}, \pi, \pi^{i}, {\tilde \pi}^{ij}\right)\;\;\;\;,
\label{cambio}
\end{equation}

which can be shown to be canonical \cite{GiontiCorfu2017}. In these coordinates, the Hamiltonian density is 

\begin{eqnarray}
{\cal H}_{ADM}&=&N\left((16\pi G)G_{abcd}{\tilde \pi}^{ab}{\tilde \pi}^{cd}-\frac{{\sqrt h}({}^{3}R-2\Lambda)}{16\pi G}\right)+2{\tilde \pi}^{ab}{\bar \nabla}_{a}N_{b}\\ \nonumber &+&
\frac{{\sqrt h}(G_{,0} -G_{,k} N^k){\bar \nabla}_{a}N^{a}}{8\pi G^2 N}
+\frac{G_{,i}{\sqrt h}h^{ij}}{8\pi G^{2}}N_{,j}\;\;\;\;,  
\label{hamiltonianagrande}  
\end{eqnarray}

$G_{abcd}$ is the DeWitt supermetric

It is straightforward to show that the primary constraint are $\pi \approx 0$ and $\pi_{i} \approx 0$. The Hamiltonian constraint $\cal{H}$ and the momentum constraint ${\cal{H}_i}$ are, respectively,  

\begin{equation}
{\cal H}=(16\pi G)G_{abcd}{\tilde \pi}^{ab}{\tilde \pi}^{cd}-\frac{{\sqrt h}({}^{3}R-2\Lambda)}{16\pi G}-\frac{{\sqrt h}(G_{,0} -G_{,k} N^k){\bar \nabla}_{a}N^{a}}{8\pi G^2 N^2}-\nabla_{j}\left(\frac{G_{,i}{\sqrt h}h^{ij}}{8\pi G^{2}}\right)
\label{vinchamiltgrande}
\end{equation}

\begin{equation}
{\cal H}_{i}=-2{\bar \nabla}^{a}{\tilde \pi}_{ai}+\frac{{\sqrt h}(-G_{,i}){\bar \nabla}_{a}N^{a}}{8\pi G^2 N}-{\sqrt h}{\bar\nabla_{i}}\left(\frac{G,_{0}-G_{,k}N^k}{8\pi G^2 N}\right)\;\;\;\;.
\label{vincconstrgrande}
\end{equation} 

The constraints analysis does not close at the secondary level. It easy to notice that the constraints are second class. At this step the invariance under diffeomorphism appears broken since we have introduced $G(x)$ and $\Lambda(x)$ which allows us to distinguish space-time points (we thank M. Reuter for this remark). In fact, for example, if we compute the variation of the three metric $h_{ij}$ and its momentum ${\tilde \pi}_{ij}$ generated by the momentum constraints ${\cal{H}}_{i}$
\begin{equation}
\{h_{ij},\int d^{3}x \tilde{N}^{i}{\cal{H}}_{i}\}={\cal{L}}_{\mathbf {\tilde{N}}} h_{ij}\;\;\;\;,
\label{tremetric}
\end{equation}

\begin{equation}
\left\{{\tilde \pi}^{ij},\int d^{3}x {\tilde N}^{i}{\cal{H}}_{i}\right\}={\cal{L}}_{\bf {\tilde{N}}} {\tilde\pi}^{ij}+{\bar \nabla}_{a}\left[\frac{{\tilde N}^{s}}{2}\left(\frac{G_{,\;s}}{8\pi G^2 N}\right)N^{a}h^{ij}\sqrt{h}\right]\;\;\;\;, 
\label{momentatrespace}
\end{equation}

and it is clear that the momenta constraints are not the generators of the space symmetries. A way to get out from all these technical difficulties is to consider instead of a general ADM metric (\ref{metricADM}), an ADM metric in Gaussian normal coordinates \cite{Wiltshire}\cite{Christodoulou}

\begin{equation}
g=-N^{2}(t)dt \otimes dt+h_{ij}dx^{i} \otimes dx^{j} 
\label{normalcoordinates}\;\;\;. 
\end{equation}

Basically we are putting the shift functions $N^{i}$ to zero, that is we are doing a gauge fixing on the spatial diffeomorphisms. The Hamiltonian density $\cal{H}_{ADM}$ then reduces to 

\begin{equation} 
{\cal H}_{ADM}=N\left((16\pi G)G_{abcd}{\tilde \pi}^{ab}{\tilde \pi}^{cd}-\frac{{\sqrt h}({}^{(3)}R-2\Lambda)}{16\pi G}\right)\;\;\;\;,
\label{Hamilrido}
\end{equation}

and the Hamiltonian constraint $\cal{H}$ is just the quantity multiplied by the shift function $N$,

\begin{equation}
{\cal H}=\left((16\pi G)G_{abcd}{\tilde \pi}^{ab}{\tilde \pi}^{cd}-\frac{{\sqrt h}({}^{(3)}R-2\Lambda)}{16\pi G}\right)\;\;\;\;. 
\label{constraint}
\end{equation}

Momenta constraints are absent and the constraint algebra easily closes. 

\section{Hamiltonian analysis of Brans-Dicke theory}

The difficulties faced in the previous Hamiltonian Analysis of the Reuter-Weyer modified Einstein-Hilbert action suggest exploring the analogous Brans-Dicke theory in the "non singular case" when Brans-Dicke is equivalent to tree level effective gravity from String Theory coupled to a dilaton field \cite{GarayGraciabellido}. In this case, the field $\phi(x)$ is dynamical with respect to the Reuter-Weyer action functional, and it has a kinetic term and a potential $U(\phi)$. We also included a York-boundary term.

\begin{equation}
S=\frac{1}{4q^2}\left[\int_{M}d^{4}x\sqrt{-g}\left(\phi^{2}\;{}^{(4)}R+4g^{\mu\nu}\partial_{\mu}\phi \partial_{\nu} \phi -U(\phi)\right)+ 2\int_{\partial M} d^3x \sqrt{h}\phi^{2}K\right]\;\;\;\;.
\label{BDaction}
\end{equation}

Its ADM decomposition, following the lines of the previous Reuter-Weyer ADM decomposition, is 

\begin{eqnarray}
S_{ADM}=\int_{t \times \Sigma} dtd^{3}x N{\sqrt{h}}\Big(\phi^{2}{}^{(3)}R+\phi^{2}K_{ij}K^{ij}-\phi^{2}K^{2}-\frac{4}{N^2}(\partial_{0}\phi)^{2}\\ \nonumber
\frac{8}{N^2}N^{i}\partial_{0}\partial_{i}\phi +4\partial_{i}\phi \partial^{i}\phi-4\frac{N^{i}N^{j}}{N^2}\partial_{i}\partial_{j}\phi+\frac{4}{N}\phi \phi,_{0}K-
\frac{4}{{\sqrt{h}N}}\phi \phi,_{i}f^{i}- U(\phi)\Big)\;\;\;.
\end{eqnarray}

We can now compute the momenta associated to $N$, $N^{i}$, $h^{ij}$ and $\phi$, and so we get 

\begin{equation}
\pi_{N}\equiv \frac{\partial {\cal L}_{ADM}}{\partial \dot{N}}\approx 0
\label{molapse}
\end{equation}

\begin{equation}
\pi_{N^i}=\frac{\partial{\cal{L}}_{ADM}}{\partial \dot {N^i}}\approx 0
\label{moshifts}
\end{equation}

\begin{equation}
\pi^{ij}=\frac{\partial{\cal{L}}_{ADM}}{\partial \dot{h}_{ij}}=-\frac{\sqrt{h}}{4q^2}\phi^2K^{ij}+\frac{1}{4}\phi\pi_\phi h^{ij}
\label{mohij}
\end{equation}

\begin{equation}
\pi_{\phi}=\frac{\partial{\cal{L}}_{ADM}}{\partial \phi}=\frac{{\sqrt{h}}}{2q^{2}N}\left(\partial_{0}\phi-N^{i}\partial_{i}\phi-\frac{N}{2}\phi K \right)
\label{momentu,phi}
\end{equation}

the first four momenta are primary constraints in close analogy to Einstein General Relativity. The total Hamiltonian $H_{T}$ is 

\begin{equation}
H_{T}=\int d^{3}x\left (\lambda_{\pi_N} \pi_{N}+ \lambda^{i}_{\pi_{N^i}}\pi_{N^i}+ N{\cal{H}}+N^{i}{\cal{H}}_{i}\right)
\label{totalHamiltonian}
\end{equation}

where the Hamiltonian Constraint $\cal H$ and the momentum constraints ${\cal H}^{i}$ are, respectively, 

\begin{equation}
{\cal{H}}=\frac{4 q^2}{\sqrt{h} \phi^2}\pi^{ij} \pi_{ij}-\frac{2 q^2}{\sqrt{h} \phi}\pi \pi_{\phi}-\frac{\sqrt{h}}{4q^2} \phi^2\; {}^{(3)}R+\frac{q^2}{2\sqrt{h}} \pi_{\phi}^2-\frac{\sqrt{h}}{4 q^2}\partial^{i}\phi\partial_{i}\phi+\frac{\sqrt{h}}{q^2}\nabla^{i}(\phi\phi,_{i})+\frac{\sqrt{h}}{4q^2}U(\phi),
\label{hamiltonianconstraint}
\end{equation}

\begin{equation}
{\cal{H}}_i=-2\nabla_j\pi^j_i+\pi_\phi\partial_i\phi\;\;\; .
\label{momentumconstraint}
\end{equation}

Here with $\pi$ we have indicated the trace of $\pi^{ij}$. Following \cite{menotti} one can show that the momentum constraints are the generators of the space diffeomorphisms on the three-dimensional surfaces  

\begin{equation}
\{h_{ij},\int d^{3}yN^l{\cal H}_l\}=\left({\cal L}_{\mathbf N} h\right)_{ij}\;\;\;\;,
\label{metricdiffeo}
\end{equation}

\begin{equation}
\{\pi^{ij},\int d^{3}yN^l{\cal H}_l\}=\left({\cal L}_{\mathbf N} \pi\right)^{ij}\;\;\;\;.
\label{momeddiffeo}
\end{equation}

These observations \cite{menotti} allow, very easily, the computation of the following commutators

\begin{equation}
\{{\cal{H}}_{i}(x), {\cal{H}}_{j}(x')\}={\cal{H}}_{i}(x')\partial_j\delta(x,x')-{\cal{H}}_{j}(x)\partial_i\delta(x,x')
\label{commumomentum}
\end{equation}

\begin{equation}
\{{\cal{H}}(x), {\cal{H}}_{j}(x')\}=-{\cal{H}}(x')\partial'_j\delta(x',x)\;\;\;\;.
\label{commuHamiltmomentum}
\end{equation}

The difficult part of this calculation is, as usual in Hamiltonian theories of General Relativity, the evaluation of the Poisson brackets between the Hamiltonian-Hamiltonian constraints. We have found the following result 

\begin{equation}
\{{\cal{H}}(x),{\cal{H}} (x')\}={\cal H}^{i}(x)\partial_{i}\delta(x,x')-{\cal H}^{i}(x')\partial'_{i}\delta(x,x')+{\chi}^{i}(x)\partial_{i}\delta(x,x')-{\chi}^{i}(x')\partial'_{i}\delta(x,x')\;\;,
\label{hamiltonian-hamiltonian}
\end{equation}

where $\chi^{i} (x)$ is defined as 

\begin{equation}
{\chi}^{i}(x)\equiv\left({\nabla}_{k}log({\phi}^{2}(x))\right)\left(8\pi^{ik}(x)-2h^{ik}\phi(x)\pi_{\phi}(x)\right)\;\;\;\;.
\label{tertiary}
\end{equation}

These Poisson brackets contain a first piece that is a linear combination of the momenta constraints as in Einstein General Relativity but also extra terms proportional to the extrinsic curvature $K^{ij}$, as we can easily see looking at \eqref{mohij}. Einstein General Relativity formulated into the Hamiltonian formalism through the ADM picture ($3+1$ decomposition) has secondary constraints, the momenta and the Hamiltonian constraints,  which are first class (see ref. \cite{menotti} \cite{espositolibro}). The algebra, through the Poisson brackets, of all Dirac's constraints of Einstein General Relativity is linear combination of the constraints \cite{menotti}, \cite{espositolibro}. Therefore we conclude  non-singular Brans-Dicke theory is, from a canonical point of view, completely different with respect to Einstein General Relativity. This feature was already highlighted by a seminal paper, in canonical analysis of Einstein General Relativity, by Hojman, Kuchar and Teitelboim \cite{Kuchar2} (we are grateful to A.Kamenshchik for pointing out this reference to us). They started from a generic scenario of a Hamiltonian geometrodynamic. Here the hypothesis is to consider a three-dimesional space-like surface $\Sigma$ embedded in the four dimensional Lorentzian manifolds $M$. $\Sigma$ evolves varing the shifts functions $N^i$ and the Lapse function $N$. Let's call these variations, respectively, $\delta N^i$ and $\delta N$. The three metric $h_{ij}$ and its conjugated momenta $\pi^{ij}$ are the sole canonical variables. The momenta constraints ${\cal{H}}^i$ and the Hamiltonian constraint ${\cal{H}}$ are functions only of the canonical variables $h_{ij}$ and $\pi^{ij}$ and they are the generators , respectively, of the variations induced by the shifts $\delta N^i$ and the lapse $\delta N$. The four constraints obey the following commutations relations

\begin{eqnarray}
\{{\cal{H}}_{i}(x), {\cal{H}}_{j}(x')\}&=&{\cal{H}}_{i}(x')\partial_j\delta(x,x')-{\cal{H}}_{j}(x)\partial_i\delta(x,x')\nonumber\\
\{{\cal{H}}(x), {\cal{H}}_{j}(x')\}&=&-{\cal{H}}(x')\partial'_j\delta(x',x)\nonumber \\ 
\{{\cal{H}}(x),{\cal{H}}(x')\}&=&{\cal H}^{i}(x)\partial_{i}\delta(x,x')-{\cal H}^{i}(x')\partial'_{i}\delta(x,x')\;\;\;\;\,
\label{algebraconstr}
\end{eqnarray}

then Einstein geometrodynamics is the only theory which satisfies all the above conditions. This results remains the same in case we consider Einstein General Relativity coupled to a field $\phi^{A}$  in a "non derivative gravitational way" \cite{Kuchar1}. This signifies that if $H^{(T)}$ is the total Hamiltonian and $H^{(M)}$ the matter part containing $\phi^{A}$, then a \textquotedblleft non derivative gravitational coupling\textquotedblright means the total Hamiltonian $H^{(T)}$ can be decomposed as a part that depends only on the geometrical variable $h_{ij}$ and $\pi^{ij}$, and another one, $H^{(M)}$, that depends by $\phi_{A}$, $\pi^{A}$ and $g_{ij}$ as  below \cite{Kuchar1}. 

\begin{equation}
H^{(T)}=H(g_{ij},\pi^{ij})+H^{(M)}(g_{ij},\phi_{A}, \pi^{A})
\label{nondervgarvit}
\end{equation}

In \cite{Kuchar2} page 131 the authors themselves recognise that the previous condition does not hold for Brans-Dicke theory.  

\section{Cosmological application to the sub-Planck Era}
 
We are now ready to apply the previous considerations to a specific Friedman Lemaitre Robertson Walker (FLRW)  minisuperspace model of RG improved Reuter-Weyer action with matter following \cite{AlessiaAlfiome} (the reader is advised to look to this reference for all the technical details). Just to fix the ideas, we start from the Einstein-Hilbert action (\ref{mEH}) now in presence of matter ${\cal{L}}_{m}$ with the York boundary term on the boundary $\partial M$ of a four dimensional Lorentian Manifold $(M,g)$

\begin{equation}
S =\int_{M} d^4 x \sqrt{-g} \, \left\{\frac{R-2\Lambda(k)}{16\pi G(k)} + \mathcal{L}_m\right\}+\frac{1}{8\pi} \int_{\partial M}{K \sqrt{h}\over G(k)}d^{3}x\;\;\;\;, 
\label{actionpunto}
\end{equation}

 and consider a FLRW metric with a lapse function $N(t)$ 

\begin{equation}
ds^2 = -N^{2}(t) dt^2 +\frac{a(t)^2}{1-K r^2} dr^2 +a(t)^2 (r^2 d\theta^2 + r^2\sin\theta d\phi^2)\;\;\;\;.
\label{FLRWmetric}
\end{equation}

 Let's suppose matter made of a barotropic perfect fluid, with density $\rho$, pressure $p$ and equation of state $p=w\rho$, $w$ being  constant. Imposing the conservation of the matter stress energy-momentum tensor $T^{\mu\nu}_{\;\;\;;\nu}=0$ we get $\rho=ma^{-3-3w}$, m being an integration constant. The matter Lagrangian density is then \cite{2014greci} ${\cal{L}}_{m}=-mNa^{-3w}$. Manrique et al. \cite{manrique} have proposed a cut off identification which is proportional to the eigenvalue of the Laplacian on the there dimensional ADM spatial-like surfaces
$k {\sim} \frac{1}{a}$ . Implementing these considerations into the Lagrangian (\ref{actionpunto}), the corresponding point Lagrangian ${\mathcal L}_{g}$ is 

\begin{equation} 
{\mathcal{L}}_g=\, -\frac{3 \, a{\dot a }^{2}}{8\pi N(t)G(a)}+\frac{3 \, a N K}{8\pi G(a)} -\frac{ a^{3}N\Lambda(a)}{8\pi G(a)}-\frac{2 Nm}{a^{3w}} +\frac{3\ , a^{2}{\dot a }^{2}G'(a)}{8\pi N G(a)^{2}}
\;\;.
 \label{lag1}
 \end{equation}
 
 We highlight  FLRW metric (\ref{FLRWmetric}) is a particular case of ADM-metric tensor in Gaussian normal coordinates \eqref{normalcoordinates}. Therefore the Hamiltonian structure should be the same described by \eqref{Hamilrido}. This is not completely true, since $G(a)$ and $\Lambda(a)$ are now functionally dependent on the dynamical variable $a(t)$ and not determined completely by the Renormalization Group. Therefore, strictly speaking, it is not a particular case of the Reuter-Weyer improved Einstein-Hilbert action. An \textquotedblleft ad hoc\textquotedblright Dirac's constraint analysis is needed. Performing it \cite{AlessiaAlfiome} \cite{GiontiCorfu2017}, a primary constraint $\pi \approx 0$, related to the lapse $N$, and a secondary constrain ${\mathcal{H}}$, that turns to be the Hamiltonian constraint, they have been found. Both are first class (see \cite{AlessiaAlfiome}) for all the technical details). The Hamiltonian constraint ${\mathcal{H}}=0$ gives the RG-improved Quantum Freedman equation
 
\begin{equation}
\frac{K}{a^2 H^2}-\frac{8\pi G(a)\, \rho+\Lambda(a)}{3 H^2}+\eta(a)+1=0 \;.
\label{QFriedEquation}
\end{equation}

The main difference of this equation with respect to the Freedman equation is the appearance of a factor $\eta(a)=-\frac{a\,G'(a)}{G(a)}$, the anomalous dimension. $K$ takes values $-1, 0, 1$ according to hyperbolic, flat, closed universes. $H=\left(\frac{\dot a}{a} \right)$ is the Hubble term. The equation of evolution for $a(t)$ becomes 

\begin{equation}
\dot{a}^2=-\tilde{V}_K(a)\equiv-\frac{K+V(a)}{\eta(a)+1}\;\;  {\textrm {where}}  \;\;
V(a)=\frac{{a}^2}{3}(8\pi G(a)\, \rho+\Lambda(a))
\label{evolvo}
\end{equation}

Notice the allowed region for dynamical evolution is $\tilde{V}_K(a)\leq 0$. Close to the non-Gaussian fixed point, using the cut off $k \sim \frac{1}{a}$, the following approximate solutions for RG-equation are deduced \cite{BonannoReuter2000}

\begin{eqnarray}
&G(a)\simeq G_0 \left(1+G_0\, g_\ast^{-1}  a^{-2}\right)^{-1} \label{grun} \nonumber \\ 
&\Lambda(a)\simeq \Lambda_0 + \lambda_\ast  a^{-2} \label{lrun} \;\;,
\end{eqnarray}

where $(G_{\ast}, \Lambda_{\ast})$ is the value of the Gravitational and Cosmological constant at the non-Gaussian fixed point. $(G_{0}, \Lambda_{0})$ are the infrared values of the Gravitation and Cosmological constants and coincide with the observed values. 

\begin{figure}
\begin{center} 
\includegraphics[width=0.55\textwidth]{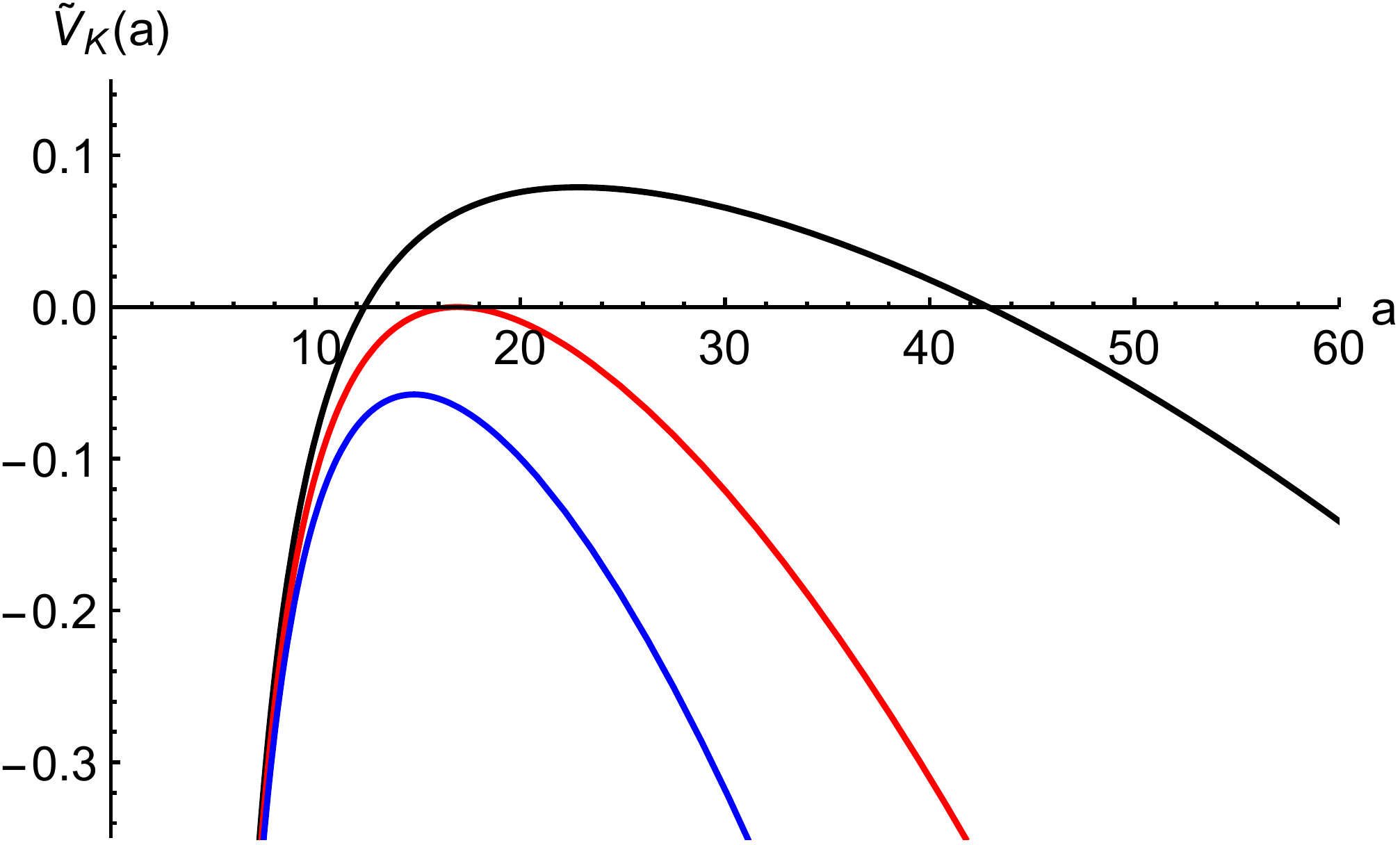}
\caption{The effective potential $\tilde{V}_K(a)$ for a
bouncing universe (black), emergent universe (red), singular universe (blue), for $K=0$, $w=1/3$, $g_\ast=0.1$, $\lambda_\ast=-0.5$ and $m=3$. Black, red and blue correspond to $\Lambda_0=2 \times 10^{-4}$, $\Lambda_0= 8.3 \times 10^{-4}$ and $\Lambda_0=1.5 \times 10^{-3}$ respectively.\label{fig1}}
\end{center}
\end{figure}

In the radiation dominated era $w=\frac{1}{3}$, the condition for Bouncing Universes $\tilde{V}_K(a)=0$ and then ${\dot {a}}(t_b)\equiv {\dot a}_b=0$ (see \cite{AlessiaAlfiome}), (\cite{GiontiCorfu2017}) for all the details) has two real solution with non negative real part

\begin{equation}
a_b^2 = -\frac{G_0 \Lambda_0 +g_\ast(\lambda_\ast-3K)}{2 g_\ast \Lambda_0} \pm\sqrt{\left(\frac{G_0 \Lambda_0-g_\ast(\lambda_\ast-3K)}{2 g_\ast \Lambda_0}\right)^2-\frac{8\pi m\,G_0}{\Lambda_0}} \;\;.
\end{equation}

The special condition for having \textquotedblleft Emergent Universes \textquotedblright \cite{ellismaartens}, \cite{AlessiaAlfiome} is ${\dot a}_b={\ddot a}_b=0$. Of course $a_b^2$ has to be positive, that is 

\begin{equation}
 \label{cond}
\lambda_\ast-3K<-\frac{G_0\Lambda_0}{g_\ast} \;\;.
\end{equation}

In Classical General Relativity $\lambda_{\ast}=0$ implies only closed universes K=1 are possible. The sub-Planckian regime in Asymptotic Safety allows, in some coupling of gravity with matter \cite{2017Alessia}, $\lambda_{\ast}$ to have a negative value so that K=0 and K=-1 are possible as well, which is a feature only of quantum era. 

We can now study  the behaviour of the early emergent universe close to $a_{b}$ linearizing the quantum equation \eqref{evolvo} around $a_{b}$. The approximate equation is then:
\begin{equation}
\dot{a}^2=\frac{4 {g_\ast} a_b^2 \Lambda_0}{3 \left({g_\ast}a_b^2-{G_0} \right)} (a-a_b)^2\;\;,
\end{equation}

then the general solution is 
\begin{equation}
a(t)=a_{b}+\epsilon\,\mathrm{exp}\left\{\sqrt{\frac{4 {g_\ast} a_b^2 \Lambda_0}{3 \left({g_\ast} a_b^2-{G_0}\right)}}\; t\right\} \;\;, 
\label{emergo}
\end{equation}

$\epsilon$ being an integration constant. It is evident that (\ref{emergo}) exibits an emergent universe scenario: there is a minimal radius $a_{b}$ and an everlasting (eternal) exponential evolution of the scale factor $a(t)$. There is no need of a model with an \textquotedblleft {\it{ad hoc}}\textquotedblright inflation. The density parameter can be written 
\begin{equation}
\Omega-1=\frac{3 \left({g_\ast}a_b^2-{G_0} \right) K}{4 {g_\ast} a_b^4 \Lambda_0}\;e^{-2N_e}\;\;.
\end{equation}
The number $N_{e}$ of e-folds is 

\begin{equation}
N_e\simeq\mathrm{log}\left(\frac{\epsilon}{a_b}\,\mathrm{exp}\left\{\sqrt{\frac{4 {g_\ast} a_b^2 \Lambda_0}{3 \left({g_\ast} a_b^2-{G_0}\right)}}\; t_e\right\}\right) \;\;,
\end{equation}
where $t_{e}$ is the cosmic time at the inflation exit. 

\section{Conclusions}

Hamiltonian analysis of the Reuter-Weyer improved Einstein-Hilbert action  has been performed. $G$ and $\Lambda$ have been treated as external non-geometrical fields. The Dirac's constraint analysis of Reuter-Weyer improved Einstein Hilbert action has been carried out up to the secondary constraint level of the Dirac's constraint analysis. It looks quite complicated and the constraints appear to be second class.  This theory behaves much better if we use \textquotedblleft Gaussian Normal Coordinates\textquotedblright. There exists a Hamiltonian constraint and it is first class. The difficulties emerged in the constraint analysis of the Reuter-Weyer  improved Einstein-Hilbert action suggest the Dirac's constraint analysis of Brans-Dicke theory. This shows a geometrodynamics with momentum constraints and Hamiltonian constraint as in Einstein General Relativity but with Dirac's constraint algebra inequivalent. In fact  the Poisson brackets of the Hamiltonian-Hamiltonian constraints are not only reducible to linear combination of constraints. Then, we conclude, the two theories, Einstein General Relativity and Brans-Dicke theory, look inequivalent as canonical (Hamiltonian) theories. 

The simple FLRW minisuperspace model based on the Reuter-Weyer improved Einstein-Hilbert action functional exibits bouncing and emergent universes for $K=0$ and $K=-1$, which are impossible to get in classical Einstein General Relativity. 

Future directions to explore are a better study of Dirac's constraint analysis of the improved Reuter-Weyer Einstein-Hilbert action functional, in particular the role of the $G(x)$ and $\Lambda (x)$ as external fields. A better understanding of the Dirac's constraint analysis of Brans-Dicke theory and the "questio disputata" of the equivalence between the Jordan frame and the Einstein frame. Finally, proceeding along these lines,  it could be also quite interesting to study ADM analysis of a Black Hole mini-supersapce models based on the improved Reuter-Weyer Einstein Hilbert action functional, which could indicate interesting features as Bouncing and Emergent Universes in the case of RG improved FLRW model in the ADM formalism.

\section{Acknowledgements}

I am very grateful to A. Bonanno for suggesting this topic of research, for discussions  and encouragment during my visits to OACT in Catania. I would like also to thank  G. Esposito, A.Kamenshchik and M. Reuter for enlightening conversations on this topic.

\section*{References}
\bibliography{castiglioncello}

\providecommand{\newblock}{}
\begin{thebibliography}{10}
\expandafter\ifx\csname url\endcsname\relax
  \def\url#1{{\tt #1}}\fi
\expandafter\ifx\csname urlprefix\endcsname\relax\def\urlprefix{URL }\fi
\providecommand{\eprint}[2][]{\url{#2}}

\bibitem{Weinberg1972}
Weinberg S 1972 {\em {Gravitation and Cosmology}\/} (New York: John Wiley and
  Sons)

\bibitem{TestGRLIGO2016}
Abbott B~P {\em et~al.\/} (LIGO Scientific, Virgo) 2016 {\em Phys. Rev.
  Lett.\/} {\bf 116} 221101 [Erratum: Phys. Rev. Lett.121,no.12,129902(2018)]
  (\textit{Preprint} \eprint{gr-qc/1602.03841})

\bibitem{Hawking1973}
Hawking S~W and Ellis G~F~R 2011 {\em {The Large Scale Structure of
  Space-Time}\/} Cambridge Monographs on Mathematical Physics (Cambridge
  University Press)

\bibitem{lauscherreuter2001}
{Lauscher} O and {Reuter} M 2002 {\em Classical and Quantum Gravity\/} {\bf 19}
  483--492 (\textit{Preprint} \eprint{hep-th/0110021})

\bibitem{wilsonfourQFT}
Wilson K~G 1973 {\em Phys. Rev.\/} {\bf D7} 2911--2926

\bibitem{NiederReuter}
{Niedermaier} M and {Reuter} M 2006 {\em Living Reviews in Relativity\/} {\bf
  9} 5

\bibitem{1979weinberg}
{Weinberg} S 1979 {\em General Relativity: An Einstein centenary survey\/} ed
  {Hawking} S~W and {Israel} W pp 790--831

\bibitem{Reuter1996}
Reuter M 1998 {\em Phys. Rev.\/} {\bf D57} 971--985 (\textit{Preprint}
  \eprint{hep-th/9605030})

\bibitem{2001ReuterSaueressig}
Reuter M and Saueressig F 2002 {\em Phys. Rev.\/} {\bf D65} 065016
  (\textit{Preprint} \eprint{hep-th/0110054})

\bibitem{Reuter2007}
Reuter M and Saueressig F 2010 {\em {Geometric and topological methods for
  quantum field theory}\/} pp 288--329 (\textit{Preprint}
  \eprint{hep-th/0708.1317})

\bibitem{WetterichFRG}
Berges J, Tetradis N and Wetterich C 2002 {\em Phys. Rept.\/} {\bf 363}
  223--386 (\textit{Preprint} \eprint{hep-ph/0005122})

\bibitem{ReuterWeyer2004}
Reuter M and Weyer H 2004 {\em Phys. Rev.\/} {\bf D69} 104022
  (\textit{Preprint} \eprint{hep-th/0311196})

\bibitem{BransDicke}
{Brans} C and {Dicke} R~H 1961 {\em Physical Review\/} {\bf 124} 925--935

\bibitem{Olmo}
Olmo G~J and Sanchis-Alepuz H 2011 {\em Phys. Rev.\/} {\bf D83} 104036
  (\textit{Preprint} \eprint{gr-qc/1101.3403})

\bibitem{ArnowittDeserMisner}
Arnowitt R, Deser S and Misner C~W 1960 {\em Phys. Rev.\/} {\bf 117}(6)
  1595--1602 \urlprefix\url{https://link.aps.org/doi/10.1103/PhysRev.117.1595}

\bibitem{1986York}
{York} J~W 1986 {\em Foundations of Physics\/} {\bf 16} 249--257

\bibitem{BonannoEspositoRubano}
{Bonanno} A, {Esposito} G and {Rubano} C 2004 {\em Classical and Quantum
  Gravity\/} {\bf 21} 5005--5016 (\textit{Preprint} \eprint{gr-qc/0403115})

\bibitem{DeWittI1967}
DeWitt B~S 1967 {\em Phys. Rev.\/} {\bf 160} 1113--1148 [3,93(1987)]

\bibitem{GiontiCorfu2017}
Gionti S~J~G 2018 {\em PoS\/} {\bf CORFU2017} 192 (\textit{Preprint}
  \eprint{gr-qc/1805.02318})

\bibitem{Wiltshire}
Wiltshire D~L 1995 {\em {Cosmology: The Physics of the Universe. Proceedings,
  8th Physics Summer School, Canberra, Australia, Jan 16-Feb 3, 1995}\/} pp
  473--531 (\textit{Preprint} \eprint{gr-qc/0101003})

\bibitem{Christodoulou}
{Christodoulou} D and {Klainerman} S 1993 {\em {The global nonlinear stability
  of the Minkowski space}\/} Princeton mathematical series ; 41 (Princeton
  University Press)

\bibitem{GarayGraciabellido}
Garay L~J and Garcia-Bellido J 1993 {\em Nucl. Phys.\/} {\bf B400} 416--434
  (\textit{Preprint} \eprint{gr-qc/9209015})

\bibitem{menotti}
{Menotti} P 2017 {\em ArXiv e-prints\/} (\textit{Preprint}
  \eprint{gr-qc/1703.05155})

\bibitem{espositolibro}
Esposito G 1992 {\em Lect. Notes Phys. Monogr.\/} {\bf 12} 1--326

\bibitem{Kuchar2}
{Hojman} S~A, {Kucha{\v r}} K and {Teitelboim} C 1976 {\em Annals of Physics\/}
  {\bf 96} 88--135

\bibitem{Kuchar1}
{Kucha{\v r}} K 1974 {\em Journal of Mathematical Physics\/} {\bf 15} 708--715

\bibitem{AlessiaAlfiome}
{Bonanno} A, {Gionti} S~J~G and {Platania} A 2018 {\em Classical and Quantum
  Gravity\/} {\bf 35} 065004 (\textit{Preprint} \eprint{gr-qc/1710.06317})

\bibitem{2014greci}
{Dimakis} N, {Christodoulakis} T and {Terzis} P~A 2014 {\em Journal of Geometry
  and Physics\/} {\bf 77} 97--112 (\textit{Preprint} \eprint{gr-qc/1311.4358})

\bibitem{manrique}
{Manrique} E, {Rechenberger} S and {Saueressig} F 2011 {\em Physical Review
  Letters\/} {\bf 106} 251302 (\textit{Preprint} \eprint{hep-th/1102.5012})

\bibitem{BonannoReuter2000}
Bonanno A and Reuter M 2000 {\em Phys. Rev.\/} {\bf D62} 043008
  (\textit{Preprint} \eprint{hep-th/0002196})

\bibitem{ellismaartens}
Ellis G~F~R and Maartens R 2004 {\em Class. Quant. Grav.\/} {\bf 21} 223--232
  (\textit{Preprint} \eprint{gr-qc/0211082})

\bibitem{2017Alessia}
Biemans J, Platania A and Saueressig F 2017 {\em JHEP\/} {\bf 05} 093
  (\textit{Preprint} \eprint{hep-th/1702.06539})

\end{thebibliography}

\end{document}